\documentclass[prb,twocolumn,showpacs,amsmath,amssymb]{revtex4}

\usepackage{color}
\usepackage{bm}
\usepackage{dcolumn}
\usepackage[dvipdfm]{graphicx}


\begin{document}

\preprint{preprint}

\title{Characterizing the weak topological properties: Berry phase point of view}

\author{Yukinori Yoshimura$^1$}
\author{Ken-Ichiro Imura$^1$}
\author{Takahiro Fukui$^2$}
\author{Yasuhiro Hatsugai$^3$}
\affiliation{$^1$Department of Quantum Matter, AdSM, Hiroshima University, Higashi-Hiroshima 739-8530, Japan}
\affiliation{$^2$Department of Physics, Ibaraki University, Mito 310-8512, Japan}
\affiliation{$^3$Institute of Physics, University of Tsukuba, 1-1-1 Tennodai, Tsukuba, Ibaraki 305-8571, Japan}

\date{\today}

\begin{abstract}

We propose classification schemes for characterizing
two-dimensional topological phases with nontrivial weak indices. 
Here, ``weak'' implies that 
the Chern number in the corresponding phase is trivial, 
while the system shows edge states along specific boundaries.
As concrete examples, we analyze different versions of the so-called Wilson-Dirac model with
(i) anisotropic Wilson terms,
(ii) next nearest neighbor hopping terms, and
(iii) a superlattice generalization of the model,
here in the tight-binding implementation.
For types (i) and (ii)
a graphic classification of {\it strong} properties 
is successfully generalized for classifying {\it weak} properties.
As for type (iii), 
weak properties are attributed
to quantized Berry phase $\pi$ along a Wilson loop.
\end{abstract}


\maketitle

\section{introduction}
The emerging field of the topological insulator has fascinated a broad perspective of
physicists, both theoretical 
\cite{QiZhang}
and experimental,
\cite{HasanKane}
because its main idea was simple
\cite{Joel}
 and the resulting topological properties are robust against disorder;
{\it i.e.}, in a sense, {\it universal}.
It has proven to be realistic and ``almost ubiquitous''.
\cite{ando}
It even exists naturally.
\cite{kawazulite}
In addition to be simple, robust and realistic,
it may also be useful.
The ``dark'' surface of the so-called ``weak'' topological insulators (WTI)
 could be etched and patterned,
in order to realize a topologically protected nanocircuit.
\cite{dark}

In contrast to the more standard ``strong'' topological insulator (STI)
that exhibits a single protected Dirac cone
(a topologically protected metallic state)
on its surface,
the WTI is called weak, because it exhibits an even number of Dirac cones
that are not necessarily protected.
A sufficient amount and a type of disorder 
may wash out characteristic features of the surface Dirac states.
\cite{KOI, liu_physica}
However,
this does not mean
that disorder has destroyed all the topological nature of the WTI phase.
Under the fragile, and damaged surface states,
topological non-triviality is simply {\it hidden};
it continues to survive,
and manifests when a proper circumstance arrives,
{\it e.g.}, when an appropriate nano-pattern is 
either formed artificially or naturally
\cite{Yazdani}
on the cleaved surface,
or in the bulk (in the from of a dislocation line),
paving the way for opening a 1D protected 
``perfectly conducting'' channel.\cite{takane}

Unfortunately, 
there have not been many experimental papers that have
reported on the nature of such WTI,
based on a study of stoichiometric compounds.
\cite{WTI_exp}
In this regard,
it would be worth mentioning
the so-called
topological crystalline insulator (TCI).
\cite{Fu_TCI, TCI_Nature2}
Unlike 
for the more standard $\mathbb{Z}_2$ topological insulator,
the spin-orbit coupling is not indispensable for TCI.
TCI is 
protected by crystalline symmetry,
\cite{Morimoto}
and exhibits an even number of Dirac cones on its surface;
{\it therefore, regarded as a variant of WTI.}
Such surface states of the TCI have been 
observed experimentally.
\cite{TCI_Nature1, TCI_Nature3}

Another context
in which
the WTI has started to be much discussed is
the superlattice of a STI and an ordinary insulator (OI).
Such a superlattice has been originally proposed
for realizing a 3D Weyl semimetal by adjusting the ratio of
the two constituent layers.
\cite{Burkov}
More recently, there are increasing number of such a superlattice
generalization of the topological insulator,
and a possibility of realizing and even {\it controlling}
(by changing the superlattice structure)
various WTI phases
\cite{Binghai, FIH, super_theo1
}
that has seemed to be hard to be realized in stoichiometric materials.

Here in this paper, we focus on 2D models, for simplicity,
and attempt to make detailed analysis and comparison
of the WTI phases
realized in superlattice and non-superlattice models.
We study variants of the so-called Wilson-Dirac model 
showing weak as well as strong topological phases. 
Here, ``strong'' means that the ground state is characterized 
by a nontrivial Chern number, 
while in the ``weak'' case the same is characterized by
a vanishing Chern number, yet shows edge states
along specific boundaries.
We consider such variations of the Wilson-Dirac model 
as with (i) anisotropic Wilson terms,
(ii) next nearest neighbor hopping terms, and also
(iii) a superlattice generalization of the model.
We show that protection of the edge states in such WTI phases 
stems from quantization of the Berry phase on an appropriate Wilson loop.
This mechanism of topological protection is much related to that of
graphene's (flat band) edge modes in the so-called ``zigzag''
edge geometry.
\cite{RyuHatsugai, Hatsugai_SSC}

The paper is organized as follows.
In Sec. II 
we introduce our models and summarize their basic properties.
A graphic way to classify weak and strong topological phases
is outlined in Sec. III.
Discussion on the superlattice model is started in Sec. IV.
In Sec. V
we study the structure of its phase diagram in detail,
and show some analytic formulas for phase boundaries.
Weak topological phases in the superlattice model 
are further analyzed from the viewpoint of quantized Berry phase
associated with a Wilson loop.
Sec. VI is devoted to conclusions.

\begin{figure*}
(a)
\includegraphics[width=80mm]{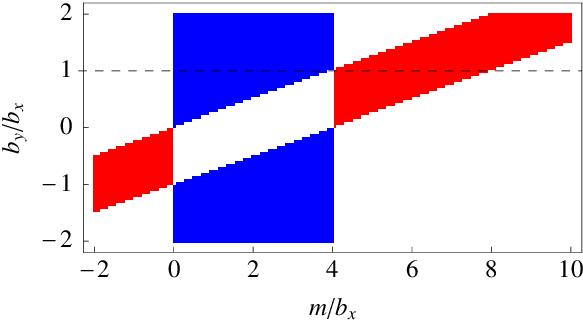}
(b)
\includegraphics[width=80mm]{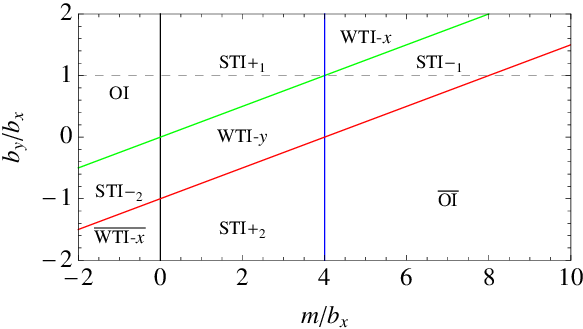}
\\
(c)
\includegraphics[width=50mm]{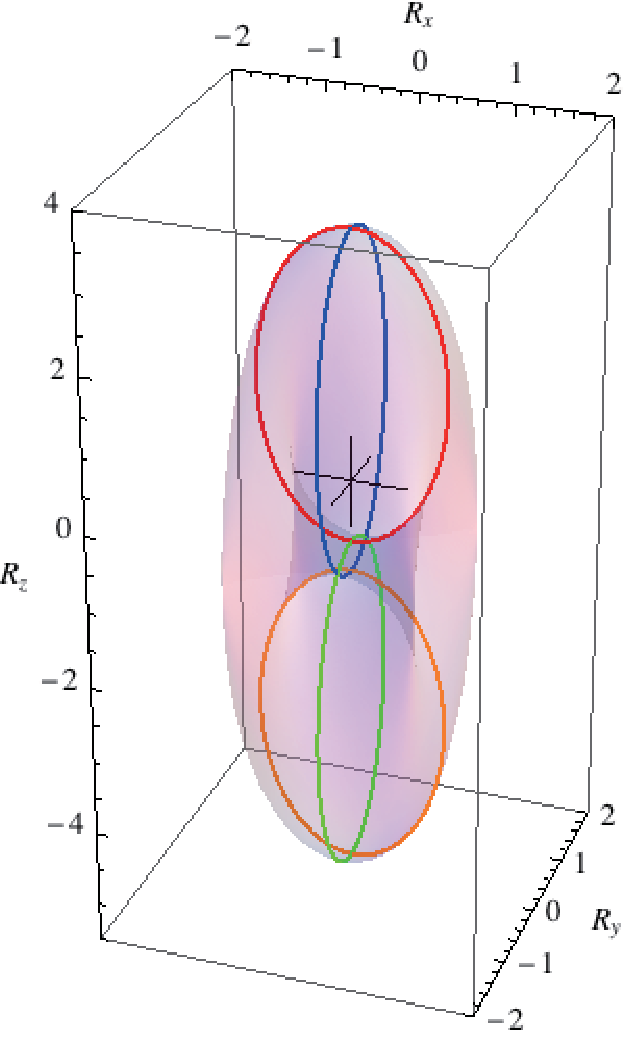}
(d)
\includegraphics[width=50mm]{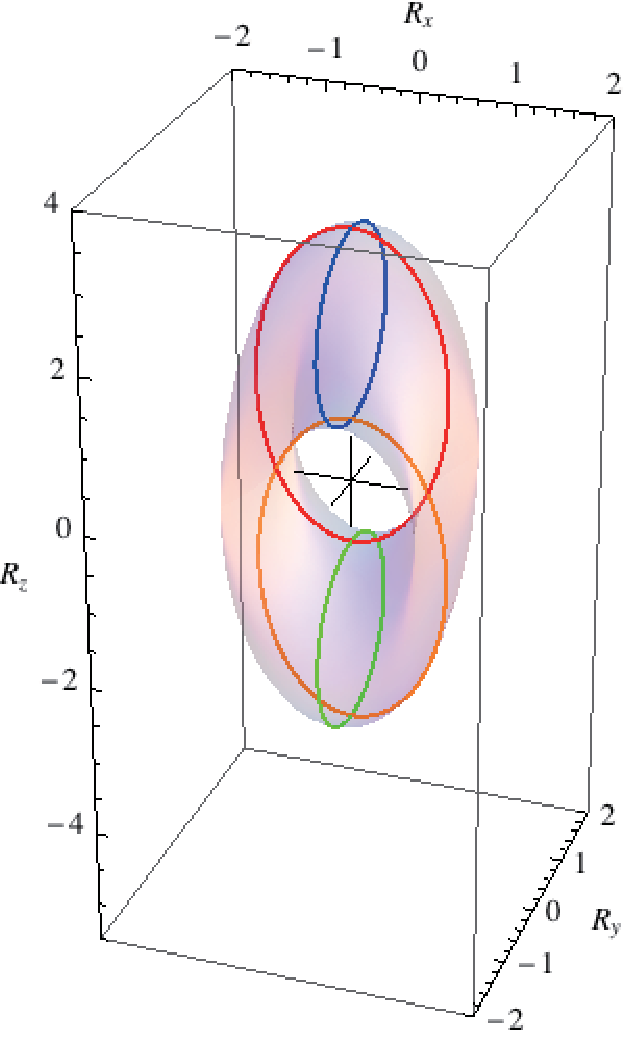}
\\
(e)
\includegraphics[width=70mm]{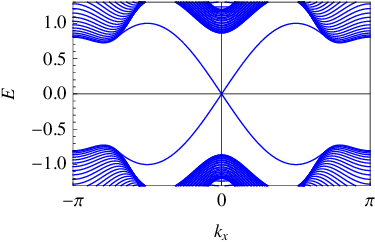}
(f)
\includegraphics[width=70mm]{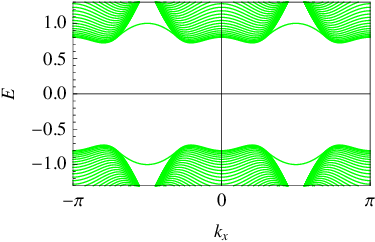}
\\
(g)
\includegraphics[width=70mm]{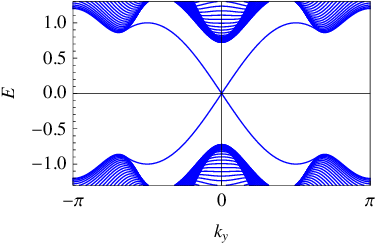}
(h)
\includegraphics[width=70mm]{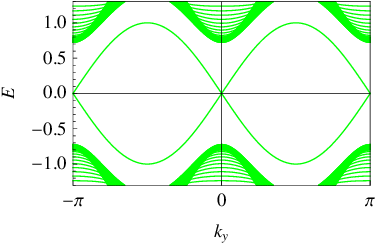}
\caption{The anisotropic Wilson-Dirac model.
(a) Phase diagram determined by numerical estimation of the bulk Chern number $\cal N$.
$b_x=1.0, t=1.0$.
(b) Phase {\it boundaries} determined by 
i) the existence vs. absence of edge modes, and
ii) closing of the bulk energy gap
at the four symmetric points:
${\bm k}=(0,0)$, $(\pi, 0)$, $(0,\pi)$ and $(\pi,\pi)$.
For each phase boundary,
we have checked that the two conditions
i) and ii) coincide.
The bulk band indices [see Eq. (\ref{FuKane})]
are also indicated in the figure.
(c), (d) Graphic representation of the closed surface ${\cal R} [T^2]$ 
and the closed loops 
${\cal R}[{\cal C}_{k_y} [0]]$(blue), ${\cal R}[{\cal C}_{k_y} [\pi]]$(green),
${\cal R}[{\cal C}_{k_x} [0]]$(red), ${\cal R}[{\cal C}_{k_x} [\pi]]$(orange)
(see Sec. III).
(e-h) Edge (and bulk) spectra
in different topological phases:
STI$+_1$ [panels (e), (g)], 
WTI-$y$ [panels (f), (h)], 
and in two different types of ribbon geometries:
$x$-oriented [panels (e), (f)], 
$y$-oriented [panels (g), (h)].}
\label{aniso}
\end{figure*}

\section{Models and their basic properties}
To realize a topologically nontrivial insulating phase on a lattice
in crystalline solids,
let us consider the following variation of the Wilson-Dirac type Hamiltonian.
\cite{BHZ, Liu_nphys, Liu_PRB}
The model is
defined on a 2D square lattice in the tight-binding approximation as
\begin{eqnarray}
H&=&
\sum_{\bm r}
\sum_{\mu=x,y}
\Big(
|\mbox{\boldmath $r$} \rangle \Gamma_{\mu} \langle \mbox{\boldmath $r+ \hat{\mu}$} |
+
| \mbox{\boldmath $r + \hat{\mu}$} \rangle \Gamma^{\dagger}_{\mu} \langle \mbox{\boldmath $r$} |
\Big)
\nonumber \\
&&+\sum_{\bm r}
|\mbox{\boldmath $r$} \rangle V(\mbox{\boldmath $r$}) \langle \mbox{\boldmath $r$}|
\label{ham1}
\end{eqnarray}
where
\begin{equation}
V({\bm r}) = \left[
m - 2 (b_x + b_y)
\right]
\sigma_z,\ 
\Gamma_\mu = -\frac{it_\mu}{2}\sigma_\mu+ b_\mu \sigma_z
\label{ham2}
\end{equation}
are, respectively, on-site potential and nearest neighbor hopping terms.
Let us  first assume that
the Wilson terms $b_\mu$
and the strength of hopping $t_\mu$
are both isotropic:
\begin{equation}
b_x = b_y\equiv b, \ 
t_x = t_y \equiv t.
\end{equation}
Then, in (crystal) momentum space,
and in the long-wavelength approximation
the Hamiltonian specified by
Eqs. (\ref{ham1}) and (\ref{ham2})
reproduces the standard Wilson-Dirac form:
\begin{equation}
h(\bm k) = t(k_x \sigma_x + k_y \sigma_y) + (m - b \bm k^2)\sigma_z.
\label{Ham2_2}
\end{equation}
Either on each site $\bm r$ in real space, or
at a given momentum $\bm k$ in the BZ, 
this model Hamiltonian 
takes a $2\times 2$ 
minimal matricial form.

On one hand,
the matrix nature of the Hamiltonian stems from 
an orbital degree of freedom
inherent to its construction.
In the presence of real spin degrees of freedom
possibly with the presence of 
time reversal symmetry
a similar minimal representation of the effective Hamiltonian 
becomes $4\times 4$.
\cite{BHZ}
On the other hand, 
if the matrix nature is due to the particle-hole symmetry of Bogoliubov-de Gennes type,
this Hamiltonian can be regarded as that for $p+ip$ superconductors.
In this case, the hopping term and the Wilson term in Eqs. (\ref{ham2}) or (\ref{Ham2_2}) become
the pairing gap and the kinetic term, respectivey. 
More rigorous equivalence is summarised in Appendix \ref{s:Chiralp}.

As demonstrated
in various contexts 
this isotropic version of the Wilson-Dirac model
exhibits only the strong topological (STI) and the ordinary (OI) 
insulating phases,
which realize in regimes:
either (i) $0<m/b<4$ or (ii) $4<m/b<8$ (for STI), and
$m/b<0$ or $8<m/b$ (for OI).
In between there is a room for realizing
a gapless Dirac semimetallic state in the clean limit.
\cite{ai_JPSJ, ai_PRB}

Since one of our purposes of the present paper is  
to study the variety of topological phases the Wilson-Dirac model has, especially
to quantify 2D
weak topological insulating (WTI) phases,
we consider in the remainder of the paper, 
three typical variation of this uniform limit.
We first allow the Wilson term $b_\mu$
to be {\it anisotropic}.
Secondly, we extends the model
to include {\it next nearest neighbor} hopping terms.\cite{Zaanen}
Finally we consider a {\it superlattice} generalization of the 
the original Hamiltonian [Eq. (\ref{ham1}) and (\ref{ham2})].

\begingroup
\renewcommand{\arraystretch}{1.5}
\begin{table}[htbp]
\caption{List of $\Delta$'s [for its definition, see Eq. (\ref{Delta}) ]
in different topological phases}
\begin{tabular}{ccccccccccccc}
\hline\hline
OI &&
$\begingroup\renewcommand\arraystretch{1.0}
\begin{bmatrix}+&-\\-&+\end{bmatrix}\endgroup$ &&
$\begingroup\renewcommand\arraystretch{1.0}
\begin{bmatrix}-&+\\+&-\end{bmatrix}\endgroup$ &&
&&
&$\begin{array}{c}\\ \\ \end{array}$&\\ \cline{2-13}
&&
OI &&
$\overline{\mbox{OI}}$ &&
&&
&&\\ \hline\hline
WTI &&
$\begingroup\renewcommand\arraystretch{1.0}
\begin{bmatrix}+&-\\+&-\end{bmatrix}\endgroup$ &&
$\begingroup\renewcommand\arraystretch{1.0}
\begin{bmatrix}-&+\\-&+\end{bmatrix}\endgroup$ &&
$\begingroup\renewcommand\arraystretch{1.0}
\begin{bmatrix}-&-\\+&+\end{bmatrix}\endgroup$ &&
&$\begin{array}{c}\\ \\ \end{array}$&\\ \cline{2-13}
&&
WTI-$x$&&
$\overline{\mbox{WTI-$x$}}$&&
WTI-$y$&&
&&
&&\\ \hline\hline
STI$_{{\cal N}=\pm 1}$ &&
$\begingroup\renewcommand\arraystretch{1.0}
\begin{bmatrix}+&-\\+&+\end{bmatrix}\endgroup$ &&
$\begingroup\renewcommand\arraystretch{1.0}
\begin{bmatrix}-&+\\+&+\end{bmatrix}\endgroup$ &&
$\begingroup\renewcommand\arraystretch{1.0}
\begin{bmatrix}-&-\\+&-\end{bmatrix}\endgroup$ &&
$\begingroup\renewcommand\arraystretch{1.0}
\begin{bmatrix}-&-\\-&+\end{bmatrix}\endgroup$ 
&$\begin{array}{c}\\ \\ \end{array}$&\\ \cline{2-13}
&&
STI$+_1$&&
STI$+_2$&&
STI$-_1$&&
STI$-_2$
&&\\ \hline\hline
STI$_{{\cal N}=\pm 2}$ &&
$\begingroup\renewcommand\arraystretch{1.0}
\begin{bmatrix}+&+\\+&+\end{bmatrix}\endgroup$ &&
$\begingroup\renewcommand\arraystretch{1.0}
\begin{bmatrix}-&-\\-&-\end{bmatrix}\endgroup$ &&
&&
&$\begin{array}{c}\\ \\ \end{array}$&\\ \cline{2-13}
&&
STI$_{{\cal N}=+2}$&&
STI$_{{\cal N}=-2}$&&
&&
&&
\\ \hline\hline
\end{tabular}
\label{FuKane}
\end{table}
\endgroup

\subsection{Anisotropic Wilson-Dirac model}
\label{s:AniWilDir}

Let us consider the case of $b_x \neq b_y$.
This leads to a visible increase 
in the diversity of topological phases [see FIG. \ref{aniso} (a)].
As shown in this phase diagram,
topological classification of the system into {\it nine} different types
is no longer simply 
due to a single combination of parameters,
as it was the case in the uniform limit;
the single combination was $m/b$, but
a function of {\it e.g.}, $m/b_x$ and $b_y /b_x$
as is the case in FIG. \ref{aniso} (a).
Two types of STI phases are specified in panel (a):
shown either in red or in blue.
They correspond to a different (nontrivial) Chern number: ${\cal N}=\pm 1$,
and can be identified by numerically estimating
this number $\cal N$.
Vanishing of the Chern number $\cal N$, on contrary, does not necessarily mean
that all of the parameter regions in this category are topologically identical and
trivial.

To uncover such topologically non-trivial phases in the ${\cal N}=0$ sector
let us introduce the following set of bulk band indices,
\cite{FuKane}
associated with four Dirac points,
${\bm k}_{D}=(0,0)$, $(\pi,0)$, $(0,\pi)$, $(\pi,\pi)$,
inherent to the model.
Table \ref{FuKane} 
lists a composition of such indices in the
different topological phases
so far identified and named
in FIG. \ref{PD} and in FIG. \ref{PB}.
Here, in our two band model,
the band indices are either $+$ or $-$,
which is defined by 
the sign of the mass terms taking into account the sign of the $\gamma$-matrices in the 
continuum limit in such a way that each index gives the Chern number $\delta_{\bm k}/2$. 
\cite{semenoff}

The band indices $\delta_{\bm k}$ can be determined in the following way.
The tight-binding Hamiltonian Eq.(\ref{ham1}) and (\ref{ham2}) can be rewritten 
in momentum space as
\begin{eqnarray}
H(\bm k)
&=&t_x \sin{k_x} \sigma_x + t_y \sin{k_y} \sigma_y\nonumber\\
&&+ (m - 2 b_x (1-\cos{k_x}) - 2 b_y (1-\cos{k_y})) \sigma_z.\nonumber\\
&&
\label{hammom}
\end{eqnarray}
At the neighborhood of the four Dirac points 
${\bm k}={\bm k}_{D}+{\bm p}$ , 
Eq. (\ref{hammom}) can be expressed as follow,
\begin{eqnarray}
H({\bm k}_{D}+{\bm p}) 
&=&\tilde{t}_x p_x \sigma_x + \tilde{t}_y p_y \sigma_y 
+ (\tilde{m} + O(|\bm p|^2)) \sigma_z,\nonumber\\
&&
\end{eqnarray}
by using ${\bm k} \cdot {\bm p}$-approximation.
The specific values of $\tilde{t_x}$, $\tilde{t_y}$ and $\tilde{m}$ for each 
symmetric points ${\bm k}_{D}$ are shown in TABLE \ref{txtym}.
Then one can define $\delta_{\bm k}$ such that
\begin{equation}
\delta_{\bm k} = {\rm sgn}[\tilde{t}_x]{\rm sgn}[\tilde{t}_y]{\rm sgn}[\tilde{m}].
\end{equation}
\begin{table}[htbp]
\caption{List of $\tilde{t}_x$, $\tilde{t}_y$ and $\tilde{m}$ 
at four Dirac points ${\bm k}_{D}$}
\begin{tabular}{ccccccccccccc}
\hline\hline
${\bm k}_{D}$&&
$\tilde{t}_x$ &&
$\tilde{t}_y$ &&
$\tilde{m}$&\\
 \hline
$(0, 0)$&&
$t_x$ &&
$t_y$ &&
$m$
\\
$(\pi, 0)$&&
$-t_x$ &&
$t_y$ &&
$m-4b_x$
\\
$(0, \pi)$&&
$t_x$ &&
$-t_y$ &&
$m-4b_y$
\\
$(\pi, \pi)$&&
$-t_x$ &&
$-t_y$ &&
$m-4b_x-4b_y$
\\
\hline\hline
\end{tabular}
\label{txtym}
\end{table}

Let us define 
the set of four indices
at ${\bm k}_{D}$ by
\begin{equation}
\Delta =
\begin{bmatrix}
\delta_{0,\pi}&
\delta_{\pi,\pi}
\\
\delta_{0,0}&
\delta_{\pi,0}
\end{bmatrix}.
\label{Delta}
\end{equation}
Then,
the four elements of $\Delta$ are related to the Chern number $\cal N$ as 
\begin{equation}
{1\over 2}\delta_{0,0} +
{1\over 2}\delta_{\pi,0} +
{1\over 2}\delta_{0,\pi}+
{1\over 2}\delta_{\pi,\pi}
=
{\cal N}.
\label{Dirac_Chern}
\end{equation}
In Eq. (\ref{Dirac_Chern})
each of the four elements
$(1/2)\delta_{\bm k}=\pm 1/2$
is a contribution from a Dirac point at ${\bm k}_D = \Gamma, X, Y, M$;
this is a standard Dirac cone argument for the Chern number.
For the Chern numbers of more generic models, see Sec. \ref{s:CheNum}, 
in which we show the details of 
the method of the numerical calculation.

Furthermore, $\Delta$
contains more information than the
single Chern number, 
and allow for full classification not only of the OI and STI,  
but also of a variety of weak TI phases.
Namely, from $\Delta$ we can know the Berry phases of the continuum Dirac fermions
at a given loop in the Brillouin zone.
It has been shown that the Berry phase is quantized as $0$ or $\pi$ for the model
with chiral symmetry, which implies, respectively, no edge state and an edge state
[more precisely, even edge states and odd edge states]
at the zero energy
if the model has boundaries.\cite{RyuHatsugai}
\cite{spectator1,spectator2}
To be more specific, another combination,
\begin{equation}
\frac{\pi}{2}\delta_{k_x,0}+
\frac{\pi}{2}\delta_{k_x,\pi}
={\cal N}_x (k_x)\pi,
\label{Nx}
\end{equation}
yielding a new quantum number (the Berry phase in unit of $\pi$) ${\cal N}_x (k_x)$
at $k_x = 0$ and at $k_x=\pi$,
specifies
whether the edge spectrum crosses (if ${\cal N}_x \bmod 2=1$) 
[or not (if ${\cal N}_x \bmod 2=0$)] 
at this momentum $k_x$,
while a different combination,
\begin{equation}
\frac{\pi}{2}\delta_{0,k_y}
+\frac{\pi}{2}\delta_{\pi, k_y}
={\cal N}_y (k_y)\pi,
\label{Ny}
\end{equation}
leads to
still another quantum number ${\cal N}_y (k_y)$
[at $k_y = 0$ and at $k_y=\pi$],
and determines
whether the edge spectrum crosses (if ${\cal N}_y \bmod 2=1$) 
[or not (if ${\cal N}_y \bmod 2=0$)] 
at the momentum $k_y$.
Eqs. (\ref{Nx}), (\ref{Ny})
can be regarded as the ``weak version'' of the
the standard Dirac cone argument, here applied to the Berry phase.
Contribution
from an isolated Dirac point to the Berry phase
is $\pm\pi/2$ 
in an appropriate gauge.
Here,
such contributions from two Dirac points 
(doublers)
on a path in the BZ torus
sum up to a quantized Berry phase $\pi$;
the value of the Wilson loop 
(the lattice version of the Berry phase)
associated with this path
(see Sec. VI for details).

As demonstrated in panel (b) of FIG. \ref{aniso},
the ${\cal N}=0$ sector of panel (a)
is indeed divided into five subregions:
OI, $\overline{\rm OI}$, WTI-$x$, $\overline{{\rm WTI-}x}$ and WTI-$y$,
in which the last three correspond to {\it weak} topological phases.
In the WTI-$x$ and WTI-$y$ phases,
two Dirac points 
($=$ crossing of the two branches of edge spectrum)
appear at $k_{1D}=0$ and at $\pi$
on the edge of a ribbon organized
in the direction specified by its name,
while the spectrum is gapped (no Dirac point)
when the ribbon is perpendicular to that direction.
In analogy with
the weak indices introduced for specifying different WTI phases in 3D,
\cite{FuKane, WTI1, WTI2, WTI3}
the above WTI-$x$ and WTI-$y$ phases may be represented, respectively, 
by analogous ``weak indices''
$(\nu_1,\nu_2)=(1,0)$, and $(\nu_1,\nu_2)=(0,1)$.

\begin{figure*}
(a)
\includegraphics[width=80mm]{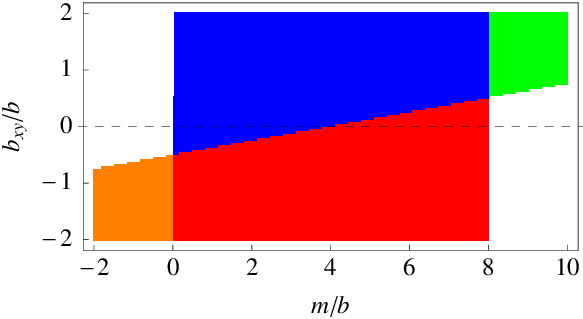}
(b)
\includegraphics[width=80mm]{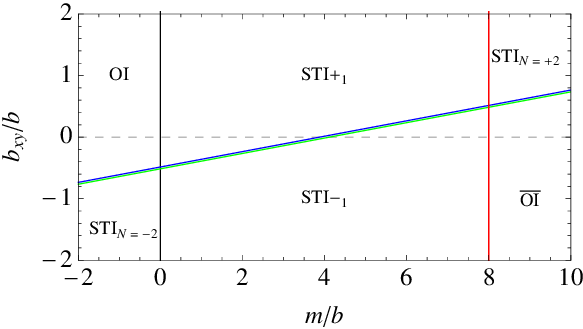}
\\
(c)
\includegraphics[width=55mm, bb=0 0 300 479]{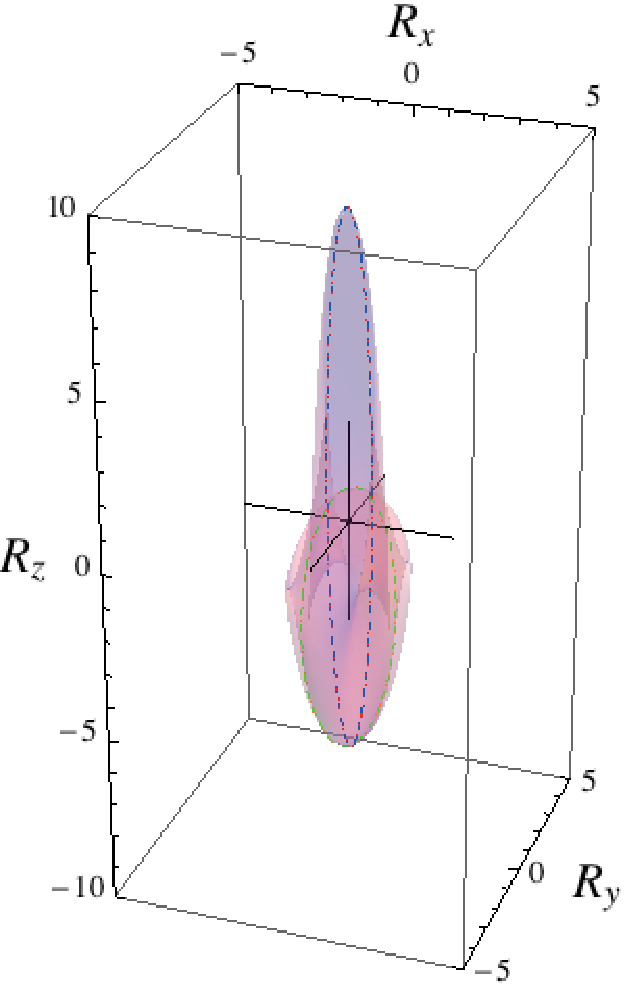}
(d)
\includegraphics[width=70mm]{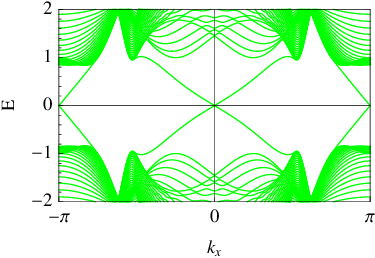}
\caption{
NNN hopping model.
(a) Phase diagram determined by numerical estimation of the Chern number $\cal N$.
Parameter regions corresponding to different ${\cal N}$ are specified by
different colors;
painted, respectively, in blue (${\cal N} =1$), red (${\cal N} =-1$), 
green (${\cal N} =2$), orange (${\cal N} =-2$), and white (${\cal N} =0$).
$b=1.0, t=1.0$.
(b) Phase boundaries, and the bulk band indices.
(c) Graphic representation of the closed surface ${\cal R} [T^2]$
and the closed loops 
${\cal R}[{\cal C}_{k_y} [0]]$(blue), ${\cal R}[{\cal C}_{k_y} [\pi]]$(green),
${\cal R}[{\cal C}_{k_x} [0]]$(red), ${\cal R}[{\cal C}_{k_x} [\pi]]$(orange)
in the STI$_{{\cal N}=2}$ phase. $m=9.0$, $b_{xy}=1.5$.
(d) The edge spectrum in the STI$_{{\cal N}=2}$ phase.
Here, the ribbon is extended along the $x$-axis, while the same spectrum is
obtained for a ribbon along the $y$-axis.
}
\label{NNN}
\end{figure*}

\subsection{Next-nearest neighbor hopping model}

Let us then consider a variation with
next-nearest neighbor (NNN) hopping:
\begin{eqnarray}
H_{NNN}&=&\sum_{\bm r}[
|{\bm r}\rangle \Gamma_1 \langle {\bm r}+\hat{\bm x}+\hat{\bm y}  |
+|{\bm r}+\hat{\bm x}+\hat{\bm y} \rangle \Gamma_1^{\dagger} \langle {\bm r} |]\nonumber\\
&&+\sum_{\bm r}[|{\bm r}\rangle \Gamma_2 \langle {\bm r}-\hat{\bm x}+\hat{\bm y}  |
+|{\bm r}-\hat{\bm x}+\hat{\bm y} \rangle \Gamma_2^{\dagger} \langle {\bm r} |]\nonumber\\
&&+\sum_{\bm r}[|{\bm r}\rangle V_{NNN}(\bm r) \langle {\bm r}|],
\label{H_NNN}
\end{eqnarray}
where
\begin{eqnarray}
\Gamma_1&=&\frac{t}{4 i}\sigma_x-\frac{t}{4i}\sigma_y+b_{xy}\sigma_z,\nonumber \\
\Gamma_2&=&\frac{t}{4 i}\sigma_x+\frac{t}{4i}\sigma_y+b_{xy}\sigma_z, \nonumber \\
V_{NNN}(\bm r)&=&-4b_{xy}\sigma_z.
\end{eqnarray}
Here, the Wilson terms are assumed to be isotropic :
$b_x =b_y =b$.
This case is of interest, since
in Ref. \cite{Zaanen}
it was proposed that the model exhibits also
weak phases, but with edge modes that appear
both in the $x$- and $y$-directions.

Different topological phases of this
NNN hoping model are shown in FIG. \ref{NNN}.
As demonstrated in the figure,
they can be characterized either by
the Chern number [see panel (a)], 
or by the bulk band indices [see panel (b)].
In contrast to the previous case of 
the anisotropic hopping model,
the NNN hopping model exhibits
strong phases with Chern number ${\cal N}= \pm 2$
[STI$_{{\cal N}=2}$ and STI$_{{\cal N}=-2}$ phases in panel (b);
see also the corresponding parameter regions in panel (a)].
In the ribbon geometry, either in the $x$- or in the $y$-direction, 
the model exhibits two Dirac cones at $k_x =0, \pi$ or at $k_y =0, \pi$ [see panel (d)].
Yet,
in contradiction with what is asserted in Ref. \cite{Zaanen},
these two Dirac cones stem from
the Chern number ${\cal N}= \pm 2$,
and the two regions should be regraded as STI phases.
In WTI phases of the anisotropic model
the two low-lying edge modes
at $k=0$ and $k=\pi$
are counter-propagating, while
here, the edge modes are co-propagating,
in consistency with
different Chern numbers in the two phases.

\subsection{The superlattice model}

The final and the most focused variation, 
as far as this paper is concerned, 
of the original model [Eq. (\ref{ham1}) and (\ref{ham2})]
is the superlattice generalization.
Here, we choose to allow the (half) band gap, the parameter $m$
that appear in definition of the on-site potential term $V$ in Eq. (\ref{ham2})
to become $\bm r$-depepndent:
$m\rightarrow m(\bm r)$;
$V$ becomes also $\bm r$-dependent:
\begin{equation}
V\rightarrow V(\bm r) = \left[
m(\bm r) - 4 b\right]
\sigma_z,
\label{V_r}
\end{equation}
and alternates such that
\begin{eqnarray}
m (\mbox{\boldmath $r$}) &=& 
\left\{
\begin{array}{l}
m_{A}\ (x \bmod q = 1, 2, \ldots, q-1)\\
m_{B}\ (x \bmod q = 0)
\end{array}
\right.
,
\label{m_r}
\end{eqnarray}
{\it i.e.}, for sake of the simplicity,
we consider the case of such patterns that vary as 
ABAB$\cdots$
($m = m_{A}, m_{B}, m_{A}, m_{B}, \cdots$ on rows $x=1, 2, 3, 4, \cdots$)
or AABAAB$\cdots$
($m = m_{A}, m_{A}, m_{B}, m_{A}, m_{A}, m_{B}, \cdots$ on rows $x=1, 2, 3, 4, 5, 6, \cdots$)
on subsequent rows on the square lattice.
The spatial profile of the change of this mass term is 
chosen to be a vertical stripe
(a one-dimensional periodic pattern,
periodic in the $x$-direction, while
translationally invariant in the $y$-direction),
modeling a semiconductor superlattice,
recently fabricated 
as a three-dimensional layered system.
\cite{super_theo1
, super_exp2}
The advantage of the superlattice construction is that
one can, in principle,
arbitrarily change and control
the profile of this pattern.

It is instructive to represent the tight-binding Hamiltonian, 
now specified by
Eqs. (\ref{ham1}), (\ref{ham2}), (\ref{V_r}) and  (\ref{m_r})
in Fourier space.
For ABAB$\cdots$
type superlattice,
this becomes
\begin{alignat}1
{\cal H}(\mbox{\boldmath $k$})=
\begin{pmatrix}
\Lambda_A & \Gamma^{}_{x} + e^{-i k_x} \Gamma^{\dagger}_{x}\\
\Gamma^{\dagger}_{x} + e^{i k_x} \Gamma^{}_{x} & \Lambda_B
\end{pmatrix},
\end{alignat}
while in the case of AABAAB$\cdots$ like pattern
the same Fourier space Hamiltonian
is represented by
\begin{alignat}1
{\cal H}(\mbox{\boldmath $k$})=
\left(
\begin{array}{ccc}
\Lambda_A  & \Gamma^{}_{x} &  e^{-i k_x} \Gamma^{\dagger}_{x}\\
\Gamma^{\dagger}_{x} & \Lambda_A  &\Gamma^{}_{x}\\
e^{i k_x} \Gamma^{}_{x} & \Gamma^{\dagger}_{x} & \Lambda_B
\end{array}
\right),
\end{alignat}
where
$\Lambda_{\alpha}=[m_{\alpha}-2b(2-\cos{k_y})] \sigma_z+t\sin{k_y}\sigma_y, (\alpha=A, B)$.
Reflecting the real space periodicity $q$
({\it e.g.}, 
$q=2$ for the ABAB, and $q=3$ for AABAAB
superlattices),
the Fourier space Hamiltonian
${\cal H} (\bm k)$ is represented by
a $q\times q$ block matrix.
The periodicity $q$ and magnitudes of 
the band gap $m_{A}, m_{B}, \cdots$
are control parameters at the level of
experimental condition
and material design.
As demonstrated in FIG. \ref{PD}
changing theses parameters, 
we can generate different topological phases.
One of our eventual purposes
would be,
by controlling them,
to switch on and off
topologically protected 
1D channels in
nanocircuits embeded 
in a system of topological insulator
superlattices.
\cite{dark}

\section{A graphic classification of weak and strong topological phases:
case of the $2\times 2$ matrix models}

In Sec. II-A and II-B,
which correspond to FIG. \ref{aniso}, FIG. \ref{NNN},
we have introduced
(i) anisotropic, and
(ii) NNN 
hopping models,
and classified their topological phases.
We have seen the following:
\begin{enumerate}
\item 
Topological classification of the system
into different Chern numbers ${\cal N}=0, \pm 1$
[${\cal N}=\pm 1$ corresponds to STI phases, while
${\cal N}=0$] corresponds either to OI or WTI phase.]
is safely applicable to these models.
\item
Further classification of the ${\cal N}=0$ sector
into OI and different WTI phases
is possible using the bulk band indices
introduced in Eq. (\ref{Delta}).
\item
In the NNN hopping model, there appear
STI$_{{\cal N}=2}$ and STI$_{{\cal N}=-2}$
phases with
a nontrivial Chern number ${\cal N}=\pm 2$.
\end{enumerate}
Here,
we first introduce a graphic representation of the 
above classification by the Chern number [point  (1) above],
introduced in Refs. \cite{manifold1, manifold2},
which allows for characterization of the {\it strong} properties,
then extends this idea 
to be applicable to characterize {\it weak} properties.

\subsection{A graphic characterization of the strong topological properties}

Let us focus on a mapping ${\cal R}$
from the BZ torus ($T^2$): a 2D space spanned by $\bm k = (k_x, k_y)$
to a closed surface ${\cal R} [T^2]$
that appears as a trajectory of
a 3D vector ${\bm R} (\bm k)$
defined in a 3D parameter space: 
$\bm R = (R_x, R_y, R_z)$.
Here, the mapping $\bm R$ encodes information
on the Wilson-Dirac Hamiltonian,
[specified by 
Eqs. (\ref{ham1}),  (\ref{ham2}), 
in case (i), while
Eqs. (\ref{ham1}),  (\ref{ham2}) and (\ref{H_NNN}), 
in case (ii)]
represented in momentum space as
\begin{equation}
h(\bm k) = \bm R (\bm k) \cdot \bm\sigma =
R_x(\bm k) \sigma_x
+R_y(\bm k) \sigma_y
+R_z(\bm k) \sigma_z.
\end{equation}
As shown in Refs. \cite{manifold1, manifold2},
information on
the bulk Chern number $\cal N$,
which is defined as
the value of
Berry curvature integrated over the 2D BZ torus,
can be transcribed,
in the 3D $\bm R$-parameter space,
into the number of times ${\cal N}_{\rm covering}$ in which
the origin in the target $\bm R$-space is
covered by the closed surface ${\cal R} [T^2]$
when
$\bm k$ sweeps once around the entire BZ torus.
Indeed, one can verify
$\cal N = - N_{\rm covering}$.
This signifies 
as shown in the two panels: (c) vs. (d) of FIG. \ref{aniso}
[parameters corresponding to (c) represent an STI, 
while those of (d) represent a WTI phase],
one can tell what the Chern number of the system is
by investigating the global behavior of 
the closed surface ${\cal R} [T^2]$ 
with respect to the origin
in the 3D $\bm R$-parameter space. 
In the STI$_{{\cal N}=2}$ region of the NNN hopping model
[see FIG. \ref{NNN} panel (c)]
the origin is covered twice by ${\cal R} [T^2]$.

\subsection{A new classification scheme for identifying WTI phases}

To extract 
the Berry phase characterizing 
the {\it weak} topological properties
from the same mapping ${\cal R}$,
let us consider a straight line ${\cal C}_{k_y} [k_x]$ 
or ${\cal C}_{k_x} [k_y]$ in the BZ,
at either $k_x$ (for ${\cal C}={\cal C}_{k_y}$)
or $k_y$ (for ${\cal C}={\cal C}_{k_x}$) fixed,
which represents a closed loop
on the BZ torus.
To characterize the weak properties in our models,
we will be specifically concerned about
how ${\cal R}$ maps these closed loops on the BZ torus;
especially,
${\cal C}_{k_y} [k_x]$ at $k_x = 0$ and $\pi$,
as well as ${\cal C}_{k_x} [k_y]$ at $k_y = 0$ and $\pi$.
Generally,
mapping of these closed loops represents also
a closed loop in the $\bm R$-space.
Note also that
these loops in the target space:
${\cal R}[{\cal C}_{k_y} [k_x=0]]$, ${\cal R}[{\cal C}_{k_y} [k_x=\pi]]$,
${\cal R}[{\cal C}_{k_x} [k_y=0]]$ and ${\cal R}[{\cal C}_{k_x} [k_y=\pi]]$
are not only on ${\cal R}[T^2]$, 
but always on a plane that includes
the origin in the $\bm R$-space. 
This implies that on the given loop, the model has chiral symmetry, and hence, 
the Berry phase should be quantized, as mentioned in Sec. \ref{s:AniWilDir}.\cite{RyuHatsugai}
If ${\cal R}[{\cal C}[k_\mu]]$ encircles the origin,
then the two branches of the corresponding edge spectrum in a ribbon geometry
laid in the $\mu$-direction
cross at this value of $k_\mu$,
while if it does not encircles the origin,
the spectrum is generally gapped at the same $k_\mu$.
Thus, 
information on the global property of
${\cal R}[T^2]$ and the four closed loops: ${\cal R}[{\cal C} [k_\mu]]$ at 
four Dirac 
points in the BZ
with respect to the origin
fully specify the weak and strong topological properties of the system.

In FIG. \ref{aniso}
panel (c) represents such an information
in a WTI-$y$ phase with ``weak indices''
$\bm \nu =(\nu_1, \nu_2) = (1,0)$
[indicating that an edge $\perp\bm\nu$ is a dark surface\cite{dark}
without surface modes]
so that both
${\cal R}[{\cal C}_{k_x}[k_y=0]]$ and ${\cal R}[{\cal C}_{k_x} [k_y=\pi]]$
encircles the origin in the $\bm R$-space,
while
neither
${\cal R}[{\cal C}_{k_y}[k_x=0]]$ nor ${\cal R}[{\cal C}_{k_y} [k_x=\pi]]$
winds the origin.

On contrary,
panel (c) in FIG. \ref{NNN}
represents the same kind of information
in the STI$_{{\cal N}=2}$ phase;
there,
all the four loops encircle the origin in the $\bm R$-space.

\begin{figure*}
(a)
\includegraphics[width=70mm]{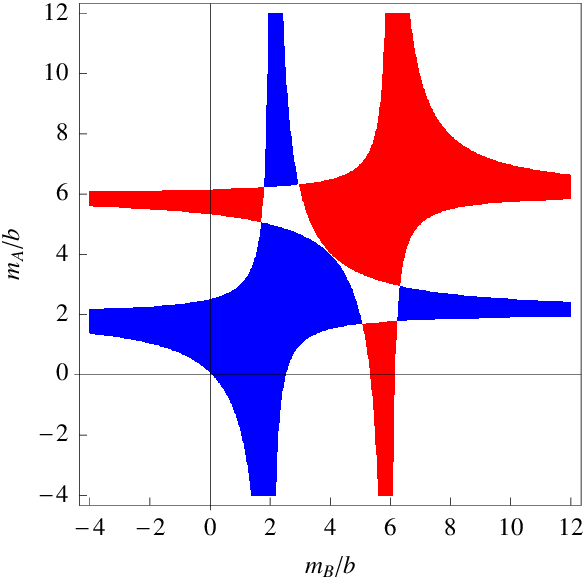}
(b)
\includegraphics[width=70mm]{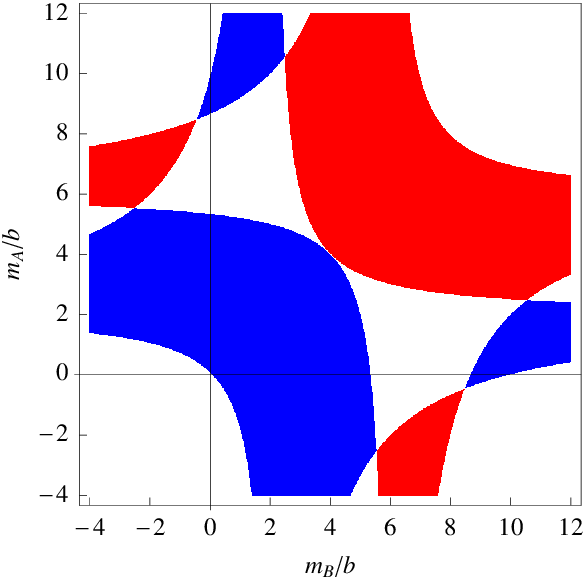}
\vspace{2mm}\\
(c)
\includegraphics[width=70mm]{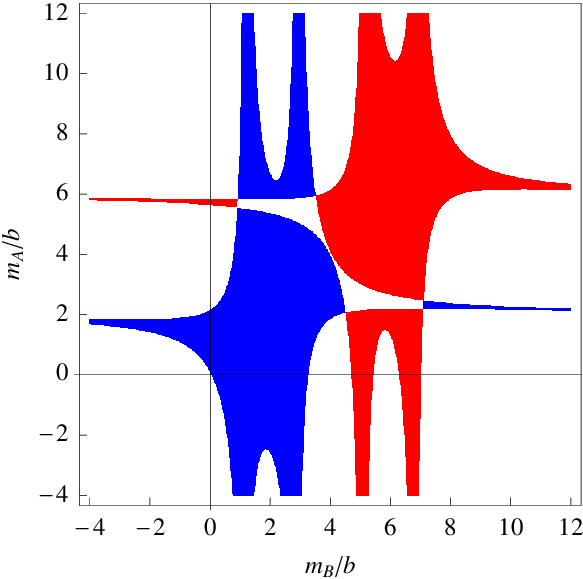}
(d)
\includegraphics[width=70mm]{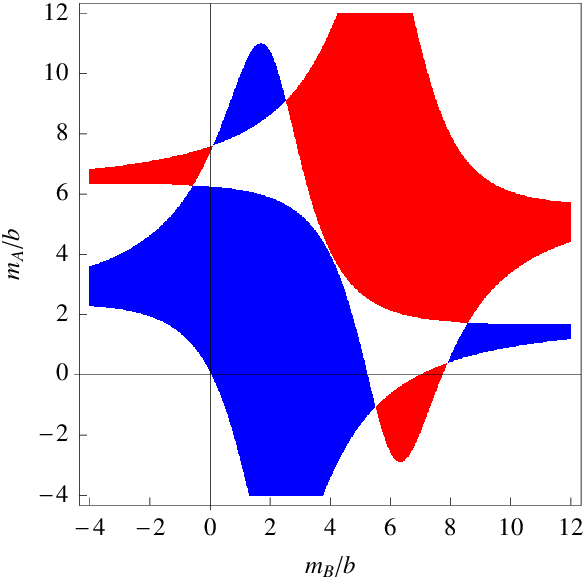}
\caption{
Phase diagram of the superlattice model
determined by numerical estimation of the Chern number ${\cal N}$.
Parameter regions represented by different colors
correspond to ${\cal N} =1$ (blue), ${\cal N} =-1$ (red), 
and ${\cal N} =0$ (white).
(a), (b): case of the ABAB$\ldots$ like pattern.
(c), (d): {\it ibid.}, AABAAB$\ldots$ type case.
$b=1.0, t=1.0$ [panels (a), (c)], $b=1.0, t=4.0$ [panels (b), (d)].}
\label{PD}
\end{figure*}
\begin{figure*}
(a)
\includegraphics[width=70mm]{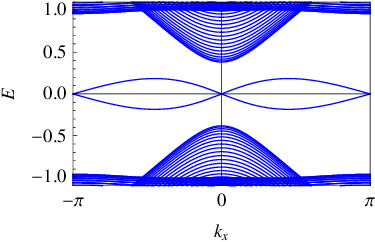}
(b)
\includegraphics[width=70mm]{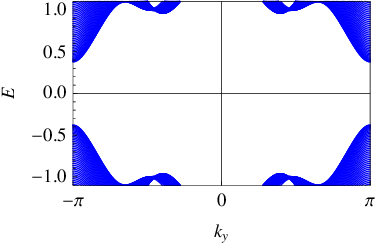}
\vspace{2mm}\\
(c)
\includegraphics[width=70mm]{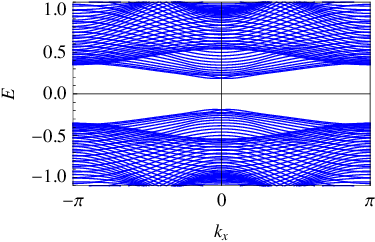}
(d)
\includegraphics[width=70mm]{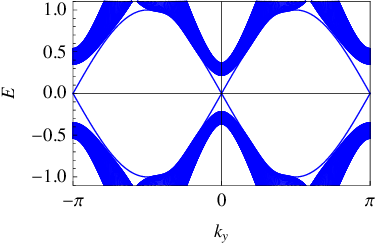}
\caption{
Energy spectrum in the ribbon geometry.
The superlattice is AABAAB $\cdots$ type.
The system is in the WTI-$x$ phase in panels (a), (b),
while it is in the WTI-$y$ phase in panels (c), (d). 
The ribbon is extended along $x$-axis in panels (a), (c),
while the same is along the $y$-axis in panels (b), (d).
}
\label{ribbon}
\end{figure*}
\begin{figure*}
(a)
\includegraphics[width=70mm]{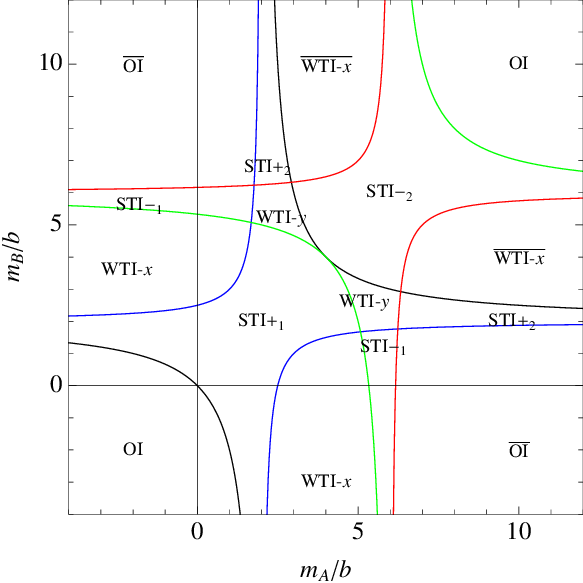}
(b)
\includegraphics[width=70mm]{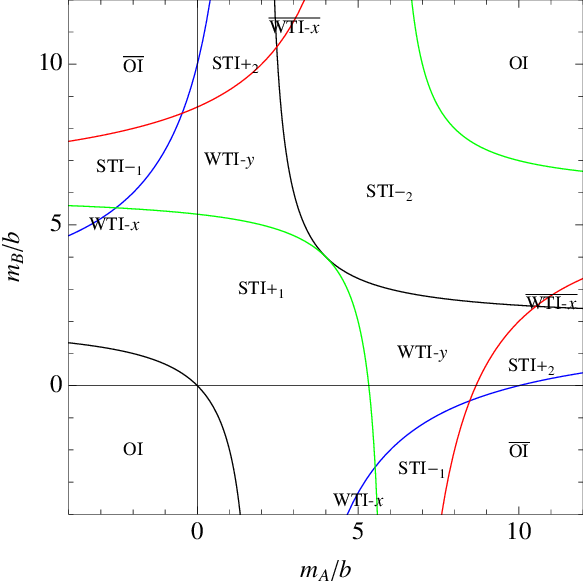}
\vspace{2mm}\\
(c)
\includegraphics[width=70mm]{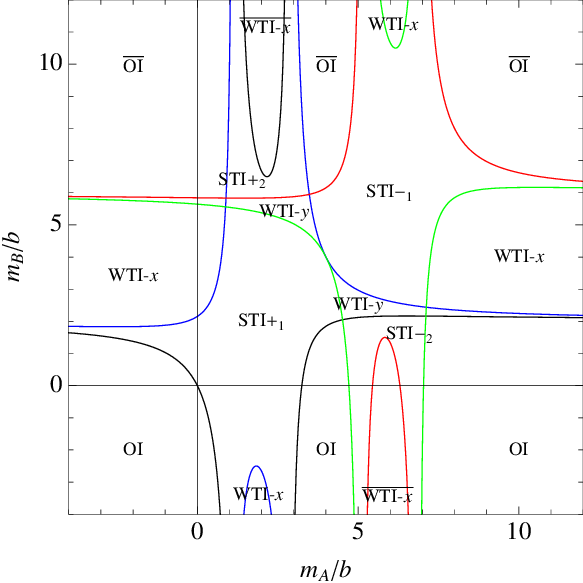}
(d)
\includegraphics[width=70mm]{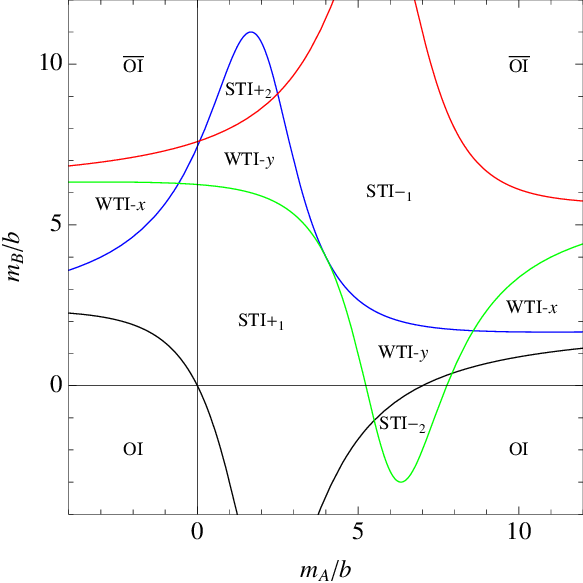}
\caption{(Color online) 
Phase boundaries between and bulk band indices in
different  topological phases.
Panels (a) and (b): cases of the ABAB$\ldots$ type pattern.
Panels (c) and (d): cases of the AABAAB$\ldots$ type pattern.
Other parameters are set the same as in the corresponding panel in FIG. \ref{PD}.
The phase boundaries are determined by tracing the zeros of
 $\det H$ at the four symmetric $\bm k$-points in the BZ (see the main text).
Different colors correspond to zeros of $\det H$ at
$(0, 0)$ (black), $(\pi, 0)$ (blue), $(0, \pi)$ (green) and $(\pi, \pi)$ (red).}
\label{PB}
\end{figure*}


\section{Phase diagram of the superlattice model}

The above classification scheme outlined in Sec. III is no longer applicable to the superlattice model,
since the superlattice structure introduces supplementary matrix nature to the model
(in addition to the original $2\times 2$ structure)
which is antithetical to the manifold scheme.
Yet, the two dimensionality of the model
allows, at least, for calculation of the Chern number
that can be used for classifying topologically nontrivial phases.
Here, applying the prescription of Ref. \cite{FukuiHatsugai}
to the superlattice system, we estimate numerically the Chern number
at different points in the $(m_A, m_B)$-plane for a given strength of hopping $t$.

\subsection{Calculation of the Chern number in multiband systems
and in the discretized BZ}
\label{s:CheNum}

For the multi-band systems, we calculate the Berry connection and curvature 
on the discretized Brillouin zone developed in Ref.\cite{FukuiHatsugai}.
On the square lattice composed of the set of discretized momentum $k_\ell$
on the Brillouin zone $[0,2\pi]\otimes[0,2\pi]$ 
defined as
\begin{alignat}1
k_\ell=\left(\frac{2\pi j_x}{L_x},\frac{2\pi j_y}{L_y}\right),
\label{DefMes}
\end{alignat}
where $j_\mu=0,1,\cdots, L_\mu$,
the Berry connection is defined by the link variable
\begin{alignat}1
U_\mu(k_\ell)=\det\psi^\dagger(k_\ell)\psi(k_\ell+\hat\mu),
\label{U_def}
\end{alignat}
where $\hat\mu$ stands for 
the unit 
lattice vector to $k_\mu$ direction,
$\hat\mu=2\pi/L_\mu$.
Manifestly gauge invariant Berry curvature is then given by the plaquette variable
\begin{alignat}1
F_{xy}(k_\ell)={\rm Im}\ln 
\left[U_x(k_\ell) U_y(k_\ell+\hat{x})U_x^{-1}(k_\ell+\hat{y})U_y^{-1}(k_\ell)\right],
\end{alignat} 
where the branch of $\ln$ is restricted to $F\in(-\pi,\pi)$.
A lattice analogue of the vortices is defined as follows.
Let $A_\mu(k_\ell)$ be a gauge potential defined by
\begin{alignat}1
A_\mu(k_\ell)={\rm Im}\ln U_\mu(k_\ell),
\end{alignat}
where the branch is also defined by $A_\mu\in(-\pi,\pi)$.
We can show
\begin{alignat}1
F_{xy}(k_\ell)=\Delta_x A_y(k_\ell)-\Delta_y A_x(k_\ell)+2\pi n_{xy}(k_\ell),
\end{alignat}
where $\Delta_\mu$ is the forward difference operator, and
$n_{xy}(k_\ell)$ is a local {\it integer} field taking $|n_{xy}|\le2$.
This $n$-field can be regarded as the vortices on the lattice.
The Chern number is given by
\begin{equation}
{\cal N}=\frac{1}{2\pi}\sum_{k_\ell}F_{xy}(k_\ell)=\sum_{k_\ell}n_{xy}(k_\ell).
\label{Chern}
\end{equation}

\subsection{Phase diagram: specific features}

To understand 
the structure of the phase diagram shown in FIG. \ref{PD}
it is convenient to first recall
what happens on the uniform line: $m_{A}=m_{B}$.
The uniform limit of the present model is nothing but
``one half'', say, the spin-up part of
the BHZ model''
\cite{BHZ}.
Also, this occurs naturally
on a diagonal line $m_{A}=m_{B}$
in phase diagrams of FIG. \ref{PD}.
In the uniform limit,
there appear three different topological 
phases,
{\it i.e.},
one trivial or ordinary (OI phase), and
two (strong) topological insulator 
(quantum anomalous Hall) phases
with a protected gapless edge mode
propagating
either in the clockwise  (STI$+$ phase in our definition) or 
in the anti-clockwise  (STI$-$ phase) direction.

Away from the uniform limit,
it is convenient to consider a phase diagram
in the $(m_{A}, m_{B})$-plane
at fixed hopping $t$.
At each point on this plane
we estimate numerically
the Chern number
to determine
which of the three topological classes
(OI, STI$+$ and STI$-$)
the system belongs to.
The resulting phase diagram is shown in
FIG. \ref{PD} [panels (a) and (d)].

In the phase diagrams shown in FIG. \ref{PD}
STI$+$ and STI$-$
phases
can be understood as a 
natural extension of
their counterparts in the uniform limit.
Yet,
as contrasted in panels
(a) vs. (b), and (c) vs. (d)
of FIG. \ref{PD}
the location of the phase boundaries
changes drastically as a function of the strength of hopping.
This is quite contrasting to the uniform limit,
and to the previous two cases treated so far,
{\it i.e.},
(i) the anisotropic hopping, and
(ii) the next-nearest neighbor hopping
cases.
Here, as shown in FIG. \ref{PD} and
in contrast to the cases (i) and (ii),
the phase boundaries in the $(m_A, m_B)$-plane
are not given by simple straight lines,
and also functions of the hopping $t$.
Naturally, the location of the phase boundaries
correspond to the closing of the bulk energy gap
triggered by the condition $\det H = 0$.
In the present realization of the superlattice pattern,
symmetries of the model
allows for closing of
the bulk energy gap at the four 
Dirac points:
$(k_x, k_y) = (0, 0)$,
$(\pi, 0)$,
$(0, \pi)$,
$(\pi, \pi)$.
Thus, 
as shown in FIG. \ref{PB}
the phase boundaries of FIG. \ref{PD}
correspond exactly to
the location of gap closing at these four symmetric points.
For this reason
the location of the phase boundaries of FIG. \ref{PD}
can be determined analytically (see Sec. V).

So far we have focused on the regions of the phase diagram
(such as the ones in FIG. \ref{PD})
painted either in blue or in red,
corresponding to regions of non-trivial Chern number
${\cal N} = \pm 1$.
Let us focus on the part in which these blue and red regions
are extended and then "overlap".
In this viewpoint
the overlap of the two nontrivial regions
({\it i.e.}, the region corresponding to
parts named ``WTI-y'' in FIG. \ref{PB}) is
a trivial region with a vanishing Chern number
${\cal N} = 0$
(unpainted, or painted in white).
It looks as if
there is a color mixture rule such that 
blue$+$red=white,
or simply the Chern numbers add to each other.
The superlattice composed of alternating stripes
with a Chern numbers
${\cal N} = 1$ and ${\cal N} = -1$,
gives rise to a phase with ${\cal N} = 0$.
However,
as implied by its name,
the WTI-$y$ phase is not actually topologically trivial,
but exhibits protected gapless edge states
in a finite geometry.
This can be explicitly verified by
numerical diagonalization of the tight-binding Hamiltonain in systems with an edge or edges.
The spectrum of such edge modes
is shown in FIG. \ref{ribbon}.
Here, we have employed
the ribbon geometry
(a system with one periodic and one open boundary condition)
so that we can distinguish
cases in which the edges are either
parallel or perpendicular to the superlattice
structure.

Naturally, WTI is an abbreviation of
the weak topological insulator.
We call this phase ``weak'',
because there are two ``Dirac cones'' on a surface,
say, in a ribbon geometry.
Here, what we mean by a Dirac cone is a crossing of
the two branches of the edge state,
one localized at the left, and the other at the right edge of the ribbon.
In the weak phase,
this crossing occurs at two points
of the now one-dimensional Brillouin zone, 
{\it e.g.} $k_y \in [-\pi,\pi]$,
and it occurs actually at the two points of the BZ,
$k_y=0$ and at $k_y= \pi$.
WTI-``$y$'', because
the protected edge states appear only in the ribbon
organized in the $y$-direction.

As demonstrated in FIG. \ref{ribbon},
this nontrivial phase
with a vanishing Chern number ${\cal N} = 0$
exhibits protected gapless states
depending on
whether
the edges are parallel or perpendicular to the superlattice.
A similar phenomenon is known
and considered to be a characteristic feature
of the more generic 3D WTI phases.

Another example
of such a nontrivial weak phase
can be found in some of the unpainted fragments
of the ${\cal N} = 0$ regions
in FIG. \ref{PD},
which are named
in FIG. \ref{PB}
the ``WTI-$x$ phase''.
As suggested in its name,
a pair of
protected edge states appear
in this phase
in a ribbon laid in the $x$-direction.

\section{Bulk-edge correspondence in the superlattice model}

We have so far seen different topological phases in the superlattice model,
and their phase boundaries
from two different points of view:
one from the Chern number in the bulk,
the other from existence vs. absence of edge states.
There is, indeed, a
one-to-one correspondence between the two viewpoints.
Existence of such a correspondence is sometimes
regarded as a defining property of the topologically nontrivial system.
Here, in this section we make more explicit the nature of
this bulk-edge correspondence in the present superlattice model.

\subsection{Bulk-edge correspondence from the viewpoint of bulk band indices}
Let us focus on the
phase boundaries between different topological phases
shown in FIG. \ref{PB}, and
marked by colored lines.
They are
colored in four colors in accordance with the number of 
the Dirac $\bm k$-points
in the (first) BZ.
Each phase boundary is a trajectory of the zeros of $\det H$
at such $\bm k$-points.
Therefore,
there are four different types of
phase boundaries,
each corresponding to a specific color
as indicated in the caption of FIG. \ref{PB}.
Then, 
comparing FIG. \ref{PB} and Table \ref{FuKane},
one can empirically verify the following 
(very important)
observation: 
{\it if one crosses such a phase boundary
corresponding to gap closing 
at a given momentum $\bm k$,
then the corresponding element of $\Delta$,
introduced in Eq. (\ref{Delta})
and listed in Table \ref{FuKane},
alters (its sign)
on the opposing sides the phase boundary.}

This observation combined with the assertions
associated with Eqs. (\ref{Nx}) and (\ref{Ny}) 
[on the (co)relation between the appearance of edge modes and
the value of quantum numbers ${\cal N}_x$ and ${\cal N}_y$]
signifies that
such a ``skeleton'' phase diagram as shown in FIG. \ref{PB},
determined simply by
the trajectory of the gap closing in the parameter space
at discrete 
Dirac points in the BZ,
contain a {\it full} topological information.
That is to say,
one can fully tell from it
whether, where and under what conditions
{\it i.e.}, in which type of the ribbon geometry
($x$-oriented or $y$-oriented),
the system exhibits protected edge modes.
In the remainder of the paper, we further analyze
and reveal
why the simple set of (band) indices 
introduced in Eq. (\ref{Delta})
acquires
such a predictive power.

\subsection{Rationalization by a transfer-matrix type argument}

Consider, for simplicity, a ribbon laid in the $x$-direction,
and introduce $\rho = e^{i k_y}$.
Focusing on one of the edges of the ribbon,
say, the one at $y=0$,
let us ask how the wave function of the edge state
penetrates from the boundary $y=0$
into the bulk region: $y>0$.
To cope with the boundary condition that
each component of the wave function vanishes,
its scalar part $\psi (y)$ is supposed to vary as
\cite{mayuko}
\begin{equation}
\psi (y) = \rho_1^y - \rho_2^y,
\label{wf}
\end{equation}
where
$\rho$'are different solutions of
the eigenvalue equation:
\begin{equation}
\det ({\cal H} - \bar{E}) =0.
\label{detH}
\end{equation}
$\bar{E}$ is the (energy) level of the edge solution,
which turns out be null to be compatible with the boundary condition
at $k_x=0$ or $\pi$
(at chirally symmetric points).
In Eq. (\ref{detH})
${\cal H}= {\cal H}(\bm k)$
is regarded as a function of $\rho$
at $k_x=0$ or $\pi$.
For the solution (\ref{wf})
to be normalizable in the region: $y>0$,
the condition $|\rho_{1,2}|<1$ must hold.

Thus, in the present context,
verifying
the correspondence between the bulk and edge properties
has reduces to establishing the equivalence of 
the following two conditions:
(i) the bulk condition:
$\det {\cal H}(k_x, 0) =0$ or $\det {\cal H}(k_x, \pi) =0$
and 
(ii) the edge condition:
$|\rho_{1,2} (k_x)|=1$,
at $k_x = 0$ and $\pi$.
Here, we verify the equivalence of two conditions
explicitly on each of the phase boundaries of FIG. \ref{PB}.
At $k_x=0$
the bulk condition is satisfied,
either by
$\det{H(0,0)}=0$
on a surface represented by
\begin{equation}
\frac{m_B}{b}=\frac{m_B \left(\frac{m_A}{b}, \frac{t}{b}\right)}{b}=
\frac{8(\frac{m_A}{b})^2-24\frac{m_A}{b}-2\frac{m_A}{b}(\frac{t}{b})^2}
{4(\frac{m_A}{b})^2-16\frac{m_A}{b}+(\frac{t}{b})^2+12},
\label{black}
\end{equation}
or by
$\det{H(0,\pi)}=0$
on a different surface:
\begin{equation}
\frac{m_B}{b}=
\frac{24 (\frac{m_A}{b})^2 -280\frac{m_A}{b} -2\frac{m_A}{b}(\frac{t}{b})^2 +12(\frac{t}{b})^2 +784}
{4(\frac{m_A}{b})^2 -48\frac{m_A}{b} +(\frac{t}{b})^2 +140}
\label{green}
\end{equation}
in the $(m_A/b, m_B/b, t/b)$-space.
On these surfaces
the edge condition $|\rho (k_x =0)|=1$ is satisfied.
Cross sections of these surfaces at $t/b=1.0$
and at $t/b=4.0$ are shown in FIG. \ref{PB} 
[cross sections of the surface (\ref{black})
are shown in black, while
those of the surface (\ref{green})
are shown in green].

Similarly, at $k_x =\pi$,
the bulk condition is satisfied either
by
$\det  H(\pi,0) =0$
on the surface:
\begin{equation}
\frac{m_{B}}{b}=
\frac{8(\frac{m_A}{b})^2 -24\frac{m_A}{b} -2\frac{m_A}{b}(\frac{t}{b})^2 +12(\frac{t}{b})^2 +16}
{4(\frac{m_A}{b})^2 -16\frac{m_A}{b} +(\frac{t}{b})^2 +12}
\label{blue}
\end{equation}
or by
$\det  H(\pi,\pi) =0$
on
\begin{equation}
\frac{m_{B}}{b}
=
\frac{24(\frac{m_A}{b})^2 -280\frac{m_A}{b} -2\frac{m_A}{b}(\frac{t}{b})^2 +24(\frac{t}{b})^2 +800}
{4(\frac{m_A}{b})^2 -48\frac{m_A}{b} +(\frac{t}{b})^2 +140}
\label{red}
\end{equation}
On these surfaces
the edge condition $|\rho (k_x =\pi)|=1$ is satisfied.
Cross sections of these surfaces at $t/b=1.0$
and at $t/b=4.0$ are shown in FIG. \ref{PB} 
[cross sections of the surface (\ref{blue})
are shown in blue, while
those of the surface (\ref{red})
are shown in red].

The fact that
(i) the bulk and (ii) the edge conditions are satisfied on the
same surfaces given by
Eqs. (\ref{black}), (\ref{green}), (\ref{blue}) and (\ref{red})
suggests that
an {\it enhanced} version of the standard bulk-edge
correspondence holds here in our system
that
involves both {\it strong} and {\it weak} topological properties.
Note that the standard bulk-edge
correspondence involves only {\it strong} properties.

\subsection{The Berry phase on a discretized Brillouin zone}

For the multi-band systems, the Berry phase is computed on the discretized Brillouin zone
(\ref{DefMes})
by the use of the link valiable defined in Eq. (\ref{U_def}) 
as the so-called Wilson loop such that
\begin{alignat}1
&W[{\cal C}_{k_y}[k_x]]=\prod_{j_y=0}^{L_y-1}U_y(k_\ell),
\nonumber\\
&W[{\cal C}_{k_x}[k_y]]=\prod_{j_x=0}^{L_x-1}U_x(k_\ell).
\end{alignat}
We take the U(1) link valuables $U_{k_\mu} (k_\ell)$ 
along a contour ${\cal C}_{k_\mu}$.
Typically, the contour ${\cal C}_{k_\mu}$ is chosen to be 
a (closed) path traversing the entire BZ,
``closed'' in the viewpoint in which
the BZ is regarded as a torus
(for the definition of contour ${\cal C}$, see Sec. III B).

The present model, as well as the original two-component Wilson-Dirac model, has 
particle-hole symmetry, which is enhanced to chiral symmetry at the four Dirac points
$\bm k=(0,0),(\pi,0),(0,\pi),(\pi,\pi)$. 
At these points, $W=1$ and $W=-1$ correspond to the Berry phase $0$ and $\pi$, respectively.

\begin{table}[htbp]
\caption{
Values of the Wilson loop ${\rm Im} \ln W [{\cal C} [k_\mu]]$ 
for specific contour ${\cal C} [k_\mu]$
at $k_\mu = 0$ and $\pi$ 
}
\begin{tabular}{cccccccccccc}
\hline
 & & ${\cal C}_{k_y} [k_x=0]$ & ${\cal C}_{k_y} [k_x=\pi]$ & & & & ${\cal C}_{k_x} [k_y=0]$ & ${\cal C}_{k_x} [k_y=\pi]$ \\ \hline\hline
STI$+_1$ &&$\pi$&$0$&&&&$\pi$&$0$\\ \hline
WTI-x &&$\pi$&$\pi$&&&&$0$&$0$\\ \hline
WTI-y &&$0$&$0$&&&&$\pi$&$\pi$\\ \hline
OI &&$0$&$0$&&&&$0$&$0$\\ 
\hline\end{tabular}
\label{VPWL}
\end{table}

\begin{figure*}[ht]
(a)
\includegraphics[width=60mm]{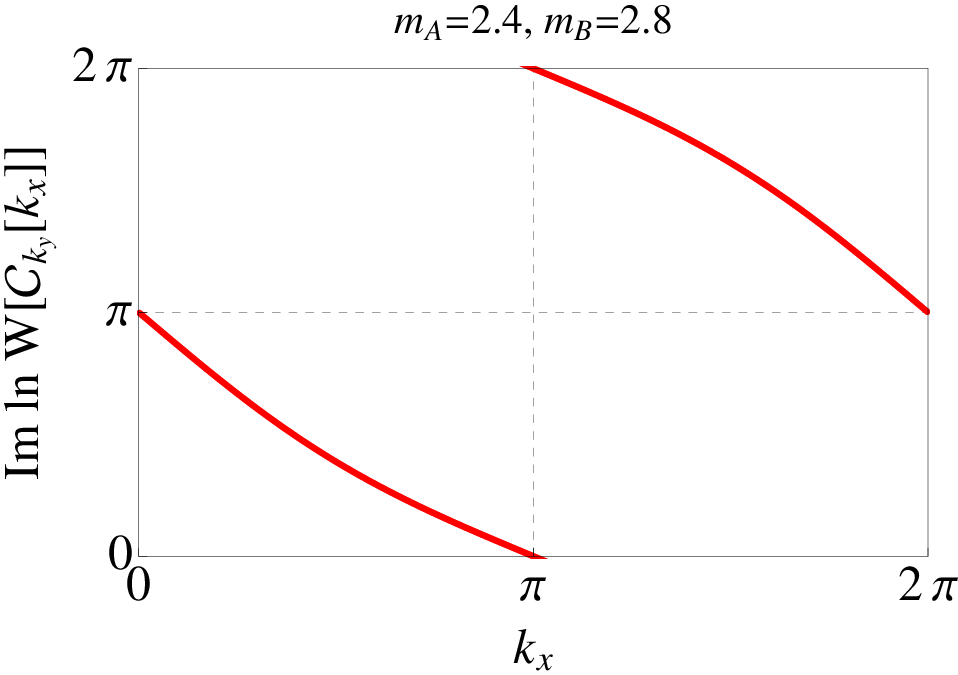}
(b)
\includegraphics[width=60mm]{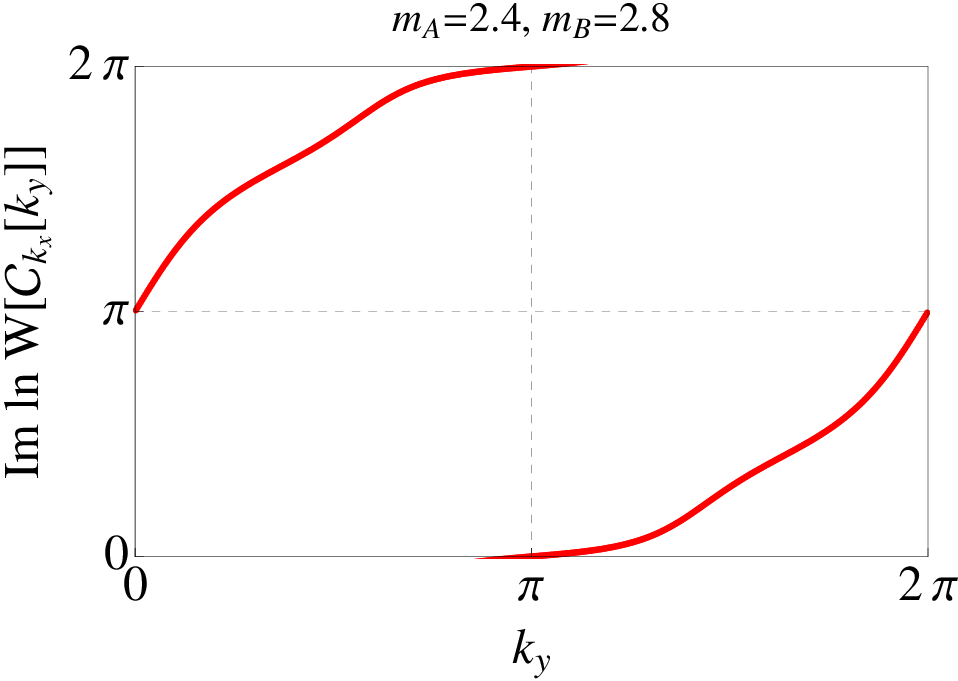}
\\
(c)
\includegraphics[width=60mm]{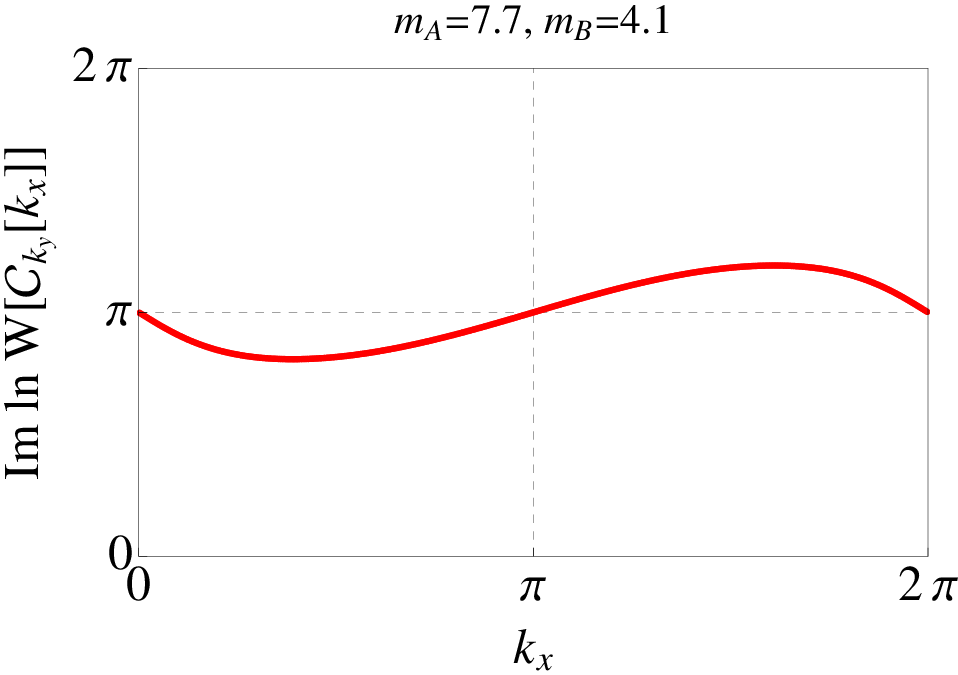}
(d)
\includegraphics[width=60mm]{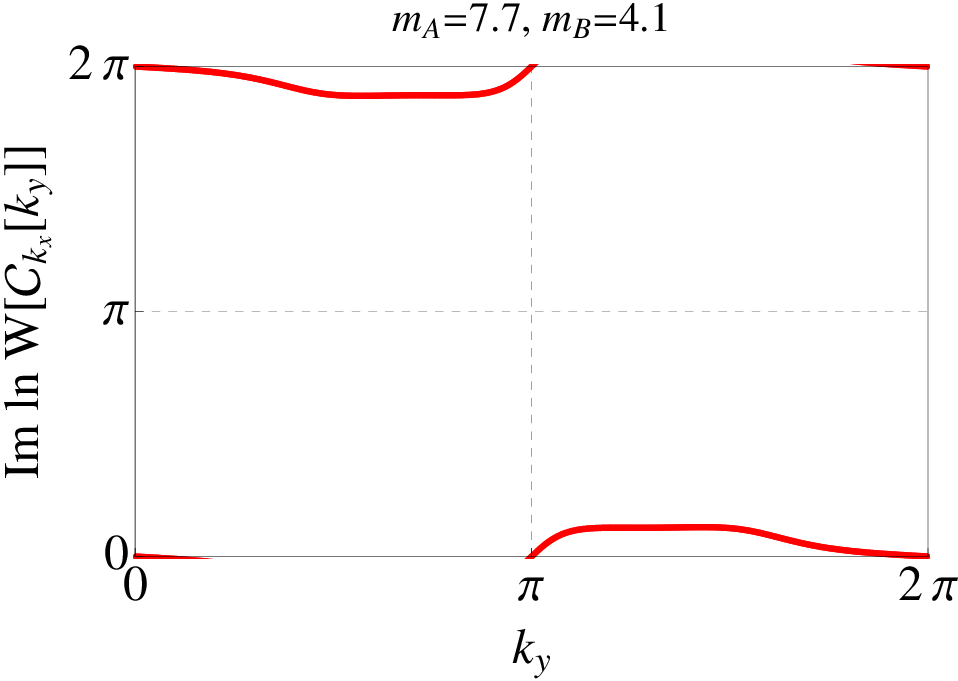}
\vspace{5mm}
\\
(e)
\includegraphics[width=60mm]{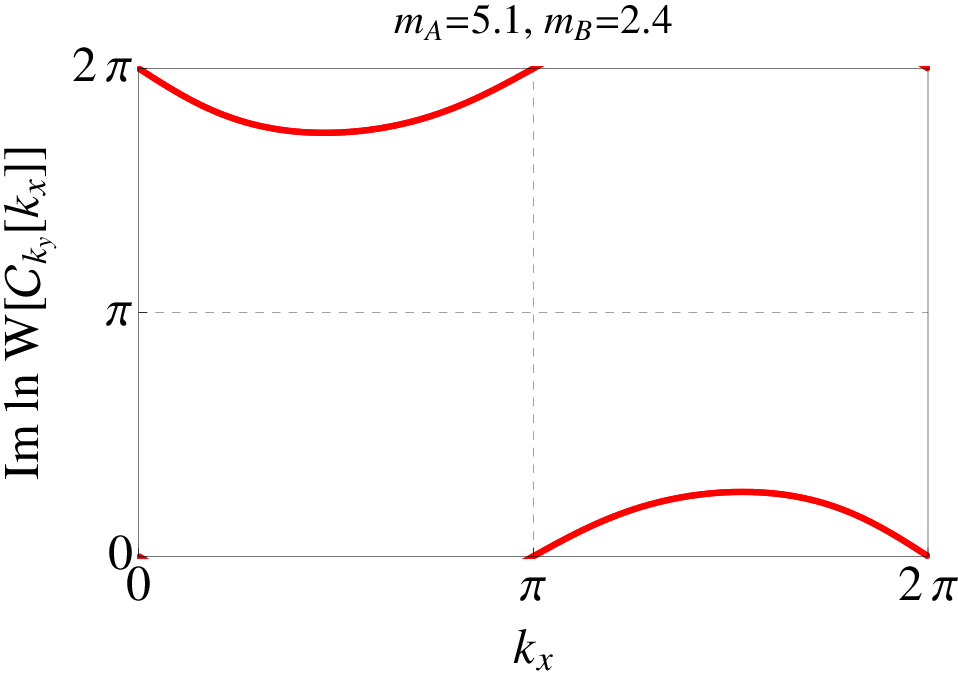}
(f)
\includegraphics[width=60mm]{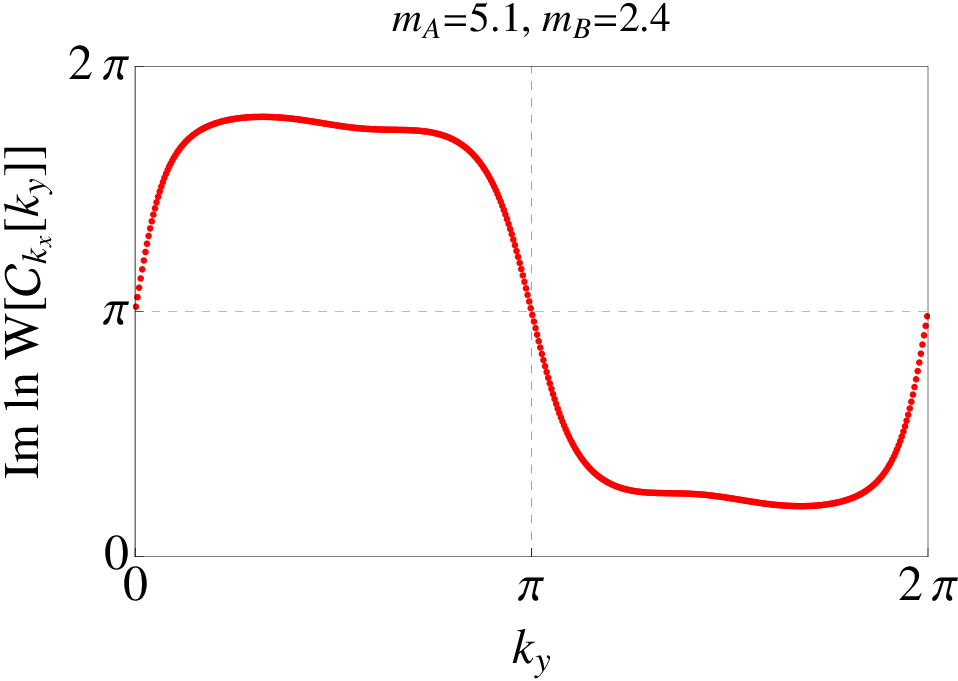}
\\
(g)
\includegraphics[width=60mm]{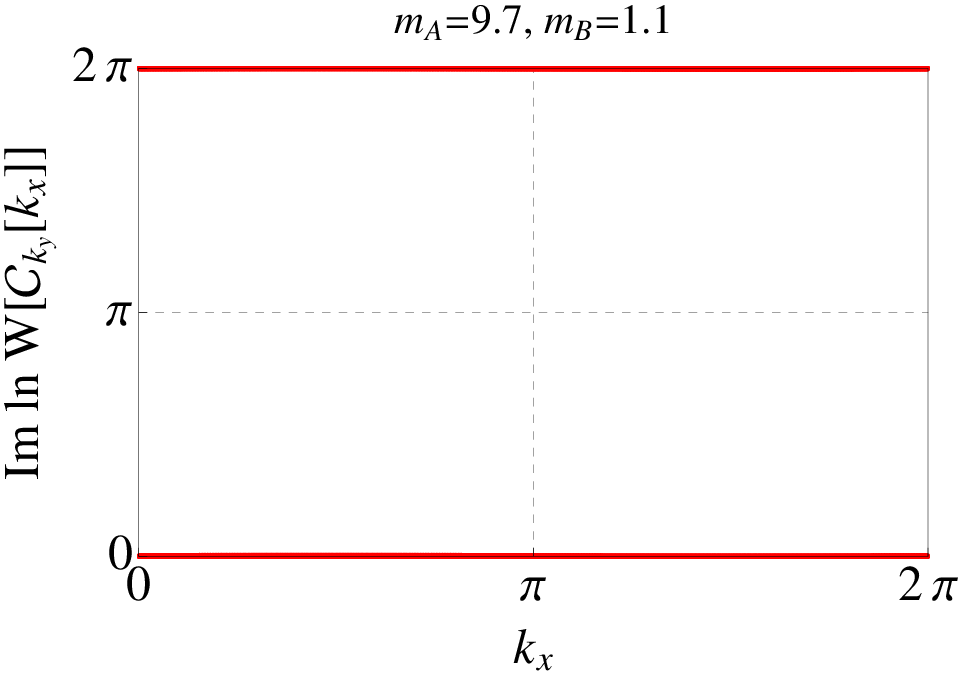}
(h)
\includegraphics[width=60mm]{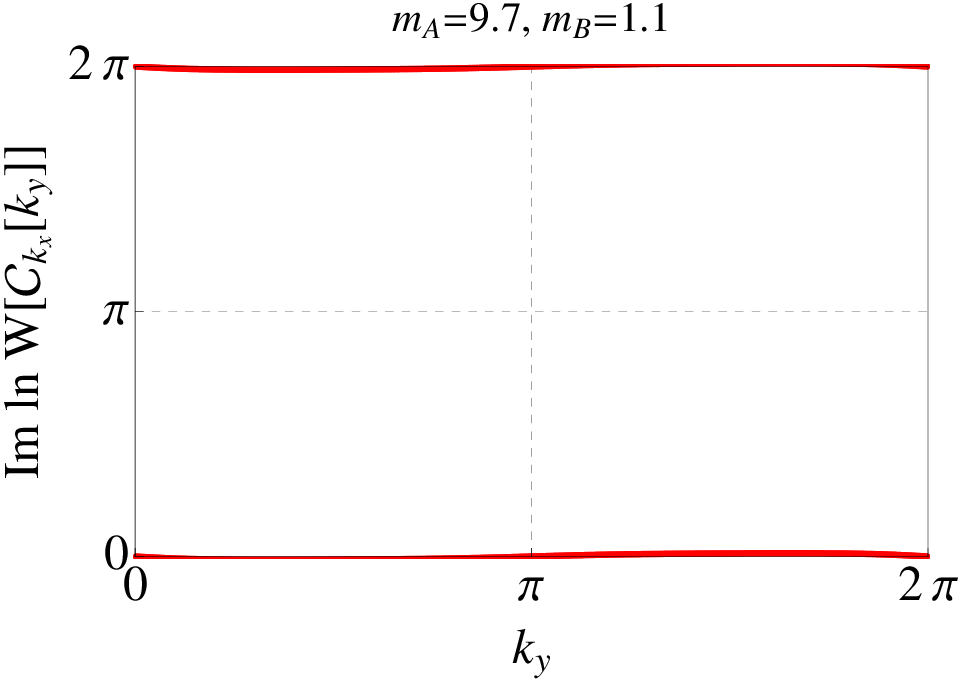}
\caption{
(Color online) Wilson loop $W[{\cal C}_{k_y}[k_x]]$ and $W[{\cal C}_{k_x}[k_y]]$.
$W[{\cal C}_{k_y}[k_x]]$ : (a), (c), (e), (g),  $W[{\cal C}_{k_x}[k_y]]$ : (b), (d), (f), (h) .
The system is $\ldots$AAB$\ldots$.
(a), (b) are case of phase STI$+_1$ (chern number=+1),
(c), (d) are case of phase WTI-$y$ (the edge state appear in the y-direction-edge) and
(e), (f) are case of phase WTI-$x$ (the edge state appear in the x-direction-edge),
(g), (h) are case of phase OI$+$.
}
\label{WL}
\end{figure*}

In FIG. \ref{WL}
typical examples of the Wilson loop calculated in the STI, WTI and OI phases are shown. 
In the left and right panels 
${\rm Im} \ln W[{\cal C}_{k_y} [k_x]]$
and ${\rm Im} \ln W[{\cal C}_{k_x} [k_y]]$
are plotted, respectively.
It should be noted that these are defined modulo $2\pi$.
The value of 
${\rm Im} \ln W[{\cal C}_{k_y} [k_x]]$
at $k_x =0,\pi$
is directly related to ${\cal N}_x (k_x)$
introduced in Eq. (\ref{Nx}),
encoding the information on the
existence of edge modes in the
$x$-oriented ribbon
at $k_x =0,\pi$,
while
${\rm Im} \ln W [{\cal C}_{k_x} [k_y]]$
at $k_y =0,\pi$
is a counterpart of ${\cal N}_y (k_y)$
given in Eq. (\ref{Ny}),
encoding the information on the
$y$-oriented ribbon
(see also Table III).
Thus,
the quantum  numbers
${\cal N}_x$ and ${\cal N}_y$
are given topological meaning:
\begin{eqnarray}
{\rm Im} \log W [{\cal C}_{k_y} [k_x]]&=& N_x (k_x)\pi + 2\pi n_x,
\\
{\rm Im} \log W [{\cal C}_{k_x} [k_y]]&=& N_y (k_y)\pi + 2\pi n_y,
\end{eqnarray}
at $k_x =0, \pi$ and $k_y =0, \pi$,
where $n_x$ and $n_y$ are arbitrary integers:
$n_x, n_y=0, \pm 1, \pm 2, \cdots$.

\section{Summary}
Motivated by recent experimental 
\cite{
 super_exp2}
realization 
and 
subsequent theoretical studies 
\cite{super_theo1, super_theo2}
of
a superlattice generalization of the topological insulator,
we have developed an extensive theoretical analysis 
of such a system,
using a model defined on a two-dimensional square lattice.
First by close diagnosis of the phase digram
composed of various types of topologically different phases,
we have established a {\it strong} correspondence
between the bulk and edge properties.
The reason why we specify that
this correspondence is ``strong''
was the following.
The word, ``bulk-edge correspondence'',
\cite{bulk-edge}
one of the key terminologies in the field of topological insulators,
signifies an idea
that (is believed to) hold in the classification of 
any type of topological non-triviality.
But in the standard use of the terminology in the field, 
it refers usually to $\bm k$-integrated topological features;
such as
the existence vs. absence
of a (pair of) protected gapless bound (localized) state(s)
at the edge (of, say, a ribbon geometry)
[in the entire spectrum; (entire) $=$
(bulk) $+$ (edge) modes, and {\it in the entire Brillouin zone} (BZ)],
in short,
whether there is an edge mode that traverses the entire bulk energy gap,
in the context of how this is related to
the (non)triviality of the bulk Chern number, or its derivatives (other topological numbers)
that is to say,
Berry curvature integrated over $\bm k$,
over the entire BZ.
The strong correspondence, here, 
addressed in this paper
deals with $\bm k$-selective information, and
refers to the so-called ``weak'' properties.
An important by-product of this analysis was the Berry phase $\pi/2$ 
that has appeared in the sum rule in Eqs. (\ref{Nx}), (\ref{Ny}).

In the second half 
of the paper we have proposed and established how
the strong correspondence of the bulk and edge properties
that plays a central role in the understanding of the
phase digram of various (weak) topological phases
is related to the {\it quantized} Berry phase along 
(Berry curvature integrated over) a Wilson loop.
In the latter terminology,
the choice of the loop (the contour of integration)
is $\bm k$-selective and allows for encoding information on the weak
topological properties.
Finally, relevance of
such ``weak'' topological properties as discussed in this paper
has also been pointed
in a slightly different context:
{\it i.e.},
in the study of network models
\cite{Mudry}
and in the context of the entanglement spectrum 
of topologically nontrivial phases.
\cite{Fakhere}

\acknowledgments
The authors acknowledge Yositake Takane, Igor Herbut,
Fakhere Assaad and Christopher Mudry 
for useful discussions.
This work was supported in part by Grant-in-Aid for Scientic Research  from
the Japan Society for the Promotion of Science (No. 25400388),
and (No. 26247064).

\appendix

\section{Equivalence to the $p+ip$ superconductors}
\label{s:Chiralp}

The Hamiltonian (\ref{ham2}) being written by creation and anihilation operators 
of the Dirac fermion is expressed as
\begin{alignat}1
H&=-\frac{it}{2}\sum_{j,\mu}\left(c_j^\dagger\gamma_\mu c_{j+\hat\mu}-h.c.\right)
\nonumber\\
&+\sum_{j,\mu}b_\mu\left(c_j^\dagger\gamma_3 c_{j+\hat\mu}+c_{j+\hat\mu}^\dagger\gamma_3c_j
-2c_j^\dagger\gamma_3c_j\right)
\nonumber\\
&+m\sum_jc_j^\dagger\gamma_3c_j ,
\end{alignat}
where $c_j=(c_{j1},c_{j2})^T$ is the anihilation operator of the Dirac fermion,
and we have chosen $\gamma_1=\sigma_x$, $\gamma_2=\sigma_z$, and $\gamma_3=-\sigma_y$.
We introduce Majorana fermion operators $a_j$ and $b_j$ such that
\begin{alignat}1
\left(\begin{array}{c} c_{j1}\\c_{j2}\end{array}\right)=
\frac{1}{2}\left(\begin{array}{c} a_{j1}+ib_{j1}\\a_{j2}+ib_{j2}\end{array}\right) ,
\end{alignat}
where $a_{j1,2}^2=b_{j1,2}^2=1$.
Then, the Hamiltonian is decoupled; $H=H_a+H_b$, where
the Majorana Hamiltonian $H_a$ is defined by\cite{EjimaFukui2011}
\begin{alignat}1
H_a&=-i\frac{t}{4}\sum_{j,\mu}a_j\gamma_\mu a_{j+\hat\mu}
+\frac{1}{2}\sum_{j,\mu}b_\mu(a_j\gamma_3a_{j+\hat\mu}-a_j\gamma_3a_j)
\nonumber\\
&+\frac{m}{4}\sum_ja_j\gamma_3a_j,
\end{alignat}
Here, $a_j=(a_{j1},a_{j2})^T$ stands for two-component Majorana operator. 
$H_b$ is the same Hamiltonian but with independent  Majorana fermon operator
$b_j=(b_{j1},b_{j2})^T$. 
Next, we introduce new Dirac fermion operators by recombining the Majorana
operators such that $\alpha_j=(a_{j1}+ia_{j2})/2$. Then, $H_a$ becomes
\begin{alignat}1
H_a&=\sum_{j,\mu}\tilde t_\mu(\alpha_j^\dagger\alpha_{j+\hat\mu}+h.c)
\nonumber\\
&+\Delta\sum_j(\alpha_j^\dagger\alpha_{j+\hat1}^\dagger-i\alpha_j^\dagger\alpha_{j+\hat2}^\dagger+h.c.)
\nonumber\\
&-\mu\sum_j(\alpha_j^\dagger\alpha_j-\alpha_j\alpha_j^\dagger) ,
\end{alignat}
where $\tilde t_\mu=-b_\mu$, $\Delta=t$, and $\mu=m/2-(b_x+b_y)$.
$H_b$ is likewise, and hence, 
it turns out that the Wilson-Dirac Hamiltonian (\ref{ham2}) also describes (two copies of)
a spinless $p+ip$ superconductor on a square lattice,
or spinful one if $H_a$ and $H_b$ are regarded as 
the Hamiltonians for spin-up and -down parts.
When $b_x=b_y\equiv b$, the chemical potential $\mu=0$ at $m=4b$, 
which is just the topological transition 
discussed by Read and Green.\cite{Read:2000nx}


\bibliography{YIFH_7}

\end{document}